# Band-Engineered LaFeO$_3$-LaNiO$_3$ Thin Film Interfaces for Electrocatalysis of Water

Rajendra Paudel[a], Andricus R. Burton[b], Marcelo A. Kuroda[a], Byron H. Farnum[b], and Ryan B. Comes[a]

[a]Department of Physics, Auburn University, Auburn, AL 36849 USA
[b]Department of Chemistry and Biochemistry, Auburn University, Auburn, AL 36849 USA

**Abstract**

Transition metal oxides have generated significant interest for their potential as catalysts for the oxygen evolution reaction (OER) in alkaline environments. Iron and nickel-based perovskite oxides have proven particularly promising, with catalytic over-potentials rivaling precious metal catalysts when the alignment of the valence band relative to the OER reaction potential is tuned through substitutional doping or alloying. Here we report that engineering of band alignment in LaFeO$_3$/LaNiO$_3$ (LFO/LNO) heterostructures via interfacial doping yields greatly enhanced catalytic performance. Using density functional theory modeling, we predict a 0.2 eV valence band offset (VBO) between metallic LNO and semiconducting LFO that significantly lowers the barrier for hole transport through LFO compared to the intrinsic material and make LFO a *p*-type semiconductor. Experimental band alignment measurements using *in situ* X-ray photoelectron spectroscopy of epitaxial LFO/LNO heterostructures agree quite well with these predictions, producing a measured VBO of 0.3(1) eV. OER catalytic measurements on the same samples in alkaline solution show an increase in catalytic current density by a factor of ~275 compared to LFO grown on *n*-type Nb-doped SrTiO$_3$. These results demonstrate the power of tuning band alignments through interfacial band engineering for improved catalytic performance of oxides.

**Introduction**

Interfaces in complex oxide thin film heterostructures are particularly interesting because of their non-equilibrium electronic properties, which usually do not exist in the bulk or even in uniform films. These unique properties emerge as a result of interfacial interactions, such as an offset in the band alignment for occupied states accompanied by charge transfer across the interface [1–12]. The band gap of the materials [13], separation between O 2p and metal 3d states [14], and film thickness [15] all play a significant role in charge transfer at the interfaces and band alignment at the surface. Careful epitaxial growth by molecular beam epitaxy or pulsed laser deposition enables control of these parameters individually and offers a route to the rational development of new functional materials for applications in renewable energy systems.

Transition metal (TM) oxide thin films, heterostructures, and interfaces have emerged as ideal model systems for understanding catalysis in energy conversion devices (electrochemical cells) by splitting of water via the oxygen evolution reaction (OER) and hydrogen evolution reaction (HER) [16–18]. The available descriptors for high OER activity are based on the number of TM 3d ($e_g$) electrons [19], the extent of O 2p-TM 3d bonding hybridization, the valence state of TM [20], and oxygen binding energy at the surface [21,22]. These parameters can be tuned by controlled doping on the perovskite A and B sites and creating interfaces and heterostructures [18,23–27]. Collectively, these effects tune the alignment of TM 3d electronic states relative to both the Fermi level and the OER reaction potential energy, which can reduce overpotentials for electrocatalysis.

LaNiO$_3$ (LNO) and other perovskite nickelates have been reported to exhibit excellent catalytic performance for OER [18,28]. Bak et al. have shown that perturbing NiO$_6$ octahedra by electrochemical exchange of Fe on the LNO surface facilitates charge transfer and improves OER activity [29]. Similar results have been reported by doping Fe on the B-site to generate partial electron transfer from Fe to Ni, leading to an Fe$^{3+\delta}$ and Ni$^{3-\delta}$ formal charge [24]. These alloy La(Fe,Ni)O$_3$ materials exhibited greater electrocatalytic performance than either pure LaFeO$_3$ (LFO) or LNO [24]. Partial oxidation of Fe$^{3+}$ results in stronger hybridization between O 2p and Fe 3d orbitals, which increases covalency resulting in higher OER activity of the perovskites [19,20]. Alternatively, forming an interface of LNO with a suitable material that induces charge transfer may produce the same effect as extrinsic doping avoiding any disorder introduced by impurity atoms, which might further improve the OER activity of LNO. Collectively, these results point to the importance of band engineering in heterostructures to boost OER activity. LaFeO$_3$ (LFO) is itself a good candidate for OER catalysis, particularly when hole doped [15,30], and given its electronic structure, LFO/LNO heterostructures might push boundaries in OER performance through interfacial hole doping of LFO. However, the LFO/LNO interface is noticeably lacking in exploration, both experimentally and theoretically.

In this work we characterize the enhanced catalytic performance of heterostructures formed with LNO/LFO grown using molecular beam epitaxy (MBE). The band alignment at LFO/LNO interface is characterized from first principles calculations using density functional theory (DFT). Taking the Zhong and Hansmann model for band alignment via continuity of the O 2p bands across an interface as a basis [14], we discuss the alignment of oxygen band center for 2p states and possible charge transfer. We also present experimental band alignment of epitaxial LFO/LNO heterostructures using X-ray photoelectron spectroscopy (XPS). Using these band-engineered heterostructures, we demonstrate the dramatic enhancement in catalytic performance of the LFO thin films by turning an intrinsic semiconductor into a *p*-type material via band engineering. These LFO/LNO heterostructures show an increase in OER catalytic reaction rate by over two orders of magnitude via cyclic voltammetry measurements compared to LFO films grown directly on *n*-doped SrTiO$_3$ [15].

**Methods**

*DFT calculations*

Computational descriptions of these heterostructures were carried out on a LaFeO$_3$/LaNiO$_3$ superlattice comprised of 4 layers of each material with a $\sqrt{2}\times\sqrt{2}$ in-plane configuration assuming periodic boundary conditions in all directions. G-type antiferromagnetic polarization for LFO [31] and no spin polarization in LNO were used to replicate the magnetic behavior of each material at room temperature [32]. In-plane lattice constants were constrained to match the lattice parameter of LFO, and equilibrium lattice parameters along out-of-plane directions and atomic coordinates were obtained by relaxation of the superlattice. Projector augmented wave (PAW) [33] pseudopotentials were used for the description of the atomic cores along with the generalized gradient approximation (GGA) of Perdew-Burke-Ernzerhof (PBE) [34] for the exchange-correlation functional. Calculations also included corrections based on the Hubbard model [35] to improve the ground state description and correct the band gap of these highly correlated electron systems [36]. The Hubbard U parameters for Fe ($U_{Fe} = 2.84$ eV), Ni ($U_{Ni} = 3.05$ eV), and O ($U_O = 6.34$ eV) reproduced the experimental band gap in the bulk LFO and LNO as previously reported in the literature [36] and were employed for heterostructure calculations. Energy cut-offs were set to 80 and 600 Ry for wave functions and charge density, respectively. The energy convergence threshold was set to $10^{-6}$ eV and integration over the superlattice Brillouin zone was performed using an 8×8×1 Monkhorst-Pack $k$-grid [37]. Our first principles calculations were performed using Quantum Espresso (QE) software suite [38].

*Synthesis and Characterization*

We synthesized epitaxial thin film heterostructures using oxide molecular beam epitaxy (MBE, Mantis Deposition). All samples were grown on 10 mm square 0.7% Nb-doped $SrTiO_3$ (STO) substrates (MTI Crystal) that served as a conductive bottom electrode for XPS and catalysis experiments. STO has a cubic lattice parameter of 3.905 Å while pseudocubic lattice parameters of LFO and LNO, obtained from experimental data are 3.93 Å [39] and 3.83 Å [32], respectively. This small lattice mismatch enables coherent growth of strained thin films. Before loading the substrates into the growth chamber, they were cleaned by sonication in acetone and iso-propyl alcohol, and then dried with molecular nitrogen.

For LNO growth, the Nb:STO substrates were heated to 600 °C in oxygen plasma generated by a 300 W-RF plasma source (Mantis Deposition) and kept at the same temperature until completion of the LNO layer. Immediately after, the substrate was heated to 750 °C for the LFO growth. Metallic fluxes were obtained by heating elemental sources in effusion cells to the evaporation or sublimation point. Elemental fluxes were calibrated using a quartz crystal microbalance (QCM). By controlling the individual shutters, LaO and $NiO_2/FeO_2$ layers were deposited alternately, as described previously [15,40]. The real-time growth was monitored using reflection high energy electron diffraction (RHEED) from the film surface. RHEED exhibited well-defined oscillations in the intensity during the entire growth, confirming the layer-by-layer growth. The oxygen pressure in the chamber during LNO growth was ~$5\times10^{-5}$ Torr with the plasma activated to maximize oxidation of the LNO film. LFO was grown at ~$6\times10^{-6}$ Torr in plasma as well. After growth, the samples were cooled in oxygen plasma to room temperature.

Chemical composition and electronic valence state of synthesized samples were established using an appended XPS system (PHI 5400, refurbished by RBD Instruments) furnished with a monochromatic Al Kα X-ray source. The absolute core level positions cannot be referenced to the Fermi level of the system owing to a electron flood gun used to mitigate charging. Thus, XPS analysis was performed by aligning O 1s peak to 530eV to provide a consistent reference for all spectra. Band alignment measurements were performed via the Kraut method using the Fe 3p and Ni 3p core level spectra, as described previously [1,41–43]. This analysis is not affected by the electron flood gun as it depends on the relative energy separation of the two core level spectra rather than the absolute binding energy as referenced to the Fermi level [43].

*OER Electrocatalysis*

To fabricate electrodes for catalytic studies, the samples were diced using a standard dicing saw to 5mm×5mm pieces. These LFO/LNO MBE films were used as electrodes and connected to a glassy carbon electrode (GC, Pine Instruments) as described in our previous work [15]. Gallium indium eutectic (InGa, Ted Pella #495425) forged electrical contact between the STO substrate and the GC working electrode. [44] Silver was used to bind the STO substrate to the surface of the GC electrode. Chemically inert epoxy (Loctite, EA E-60HP) was then used to cover any exposed silver paint and seal the edges of the STO substrate.

Cyclic voltammetry (CV) experiments were performed with a Pine WaveDriver 20 bipotentiostat using a three-electrode setup. The working, reference, and counter electrodes for the electrochemical setup were LFO/LNO, Hg/HgO (0.1 M KOH, Pine Instruments), and platinum coil respectively. All measurements were carried out under saturated $O_2$ conditions in water (18 MΩ, Millipore) with 0.1 M KOH electrolyte (pH 12.7) while rotating the working electrode at 2000 rpm to remove bubbles from the electrode surface. All potentials were converted from Hg/HgO to RHE using $[Fe(CN)_6]^{3-/4-}$ as an external standard. CV experiments were performed by sweeping the potential at 20 mV s$^{-1}$ from 0.83 to 2.23 V vs RHE for 25 cycles to equilibrate the electrode surface. The anodic trace of the 25$^{th}$ cycle was used for analysis of electrocatalytic performance for OER.

## Results

### DFT analysis

We first employ first principles calculation to establish the band alignment and charge transfer across the LFO/LNO interface and better understand how the heterostructures will function as OER catalysts. Strain and bonding environment may alter the electronic properties significantly between a bulk material and heterostructure, necessitating care to decouple interfacial effects from the structural changes imposed by the superlattice [45]. The interface was modeled employing LFO/LNO superlattice with 4 unit cell thickness for each material, as depicted in Figure 1(a). The electronic structure of the interface is rationalized by projecting Bloch states onto localized atomic orbitals in different layers of the superlattice. This decomposition, shown in Figure 1(b), only contains contributions from Fe, Ni and O orbitals as Bloch states formed with La ones only give small contributions in the vicinity of from Fermi level. The LFO bandgap in the heterostructure is estimated to be ~ 2.2 eV, which is in close agreement with the 2.3 eV experimental band gap reported previously [1,46]. Overall, features of the electronic structure are also similar to those obtained in calculations of bulk LFO and LNO presented in Figure S1 and S2 of the supplemental material. For instance, the bands crossing the Fermi level within the LNO layers are strongly hybridized Ni $e_g$ and O 2p orbital bringing metallic character to the heterostructure. Occupied Fe $e_g$ states form the top of the valence band within the LFO layers and unoccupied $t_{2g}$ states are ~2.2 eV above the LFO valence band maximum, which is consistent with the bulk properties. The LFO valence band maximum is formed of hybridized O 2p and occupied $Fe^{3+}$ $e_g$ states and lies ~0.2 eV below the Fermi level (**Figure S3**).

In general, quantitative prediction of band offsets at complex oxide interfaces using computational methods is challenging due to electron correlations, various magnetic states and symmetry considerations [47]. Nonetheless, results could be anticipated from the bulk calculations following the approach by Zhong and Hansmann [14] based on the equilibration of oxygen states yielding the energy continuity of the O 2p band center—the center of mass of the oxygen partial density of states—across the interface.

In these heterostructures, p-doping of the semiconductor is expected to be emerge as in bulk the O 2p-band center of LNO and LFO reside at approximately 2.20 eV and 2.82 eV below the Fermi level, respectively. Once the interface is formed, O 2p-band centers obtained from orbital projections on different layers tend to align, as shown in Figure 1(b) and listed in Table 1. For each material in the heterostructure we observe small differences in the calculated O 2p band centers likely attributable to local octahedral tilting and spin configuration variations across the interface that break the symmetry of the structure. The O 2p center difference for the $NiO_2$ planes in LNO reside nearly 0.15 eV above those of $FeO_2$ in LFO. Moreover, results from DFT calculations show that the valence band offset (VBO) of ~ 0.2 eV across the LNO layers, resulting in the nondegenerate p-doping of the LNO. From this offset, we estimated a hole density in the LNO to be $6.02\times10^{16}$ cm$^{-3}$ induced from the charge transfer. In further analysis of the electronic band structure, we calculated the effective mass of LNO holes of approximately 0.61$m_e$ (**Figure S4**). Details can be found in supporting information and hole densities are presented in table S1.

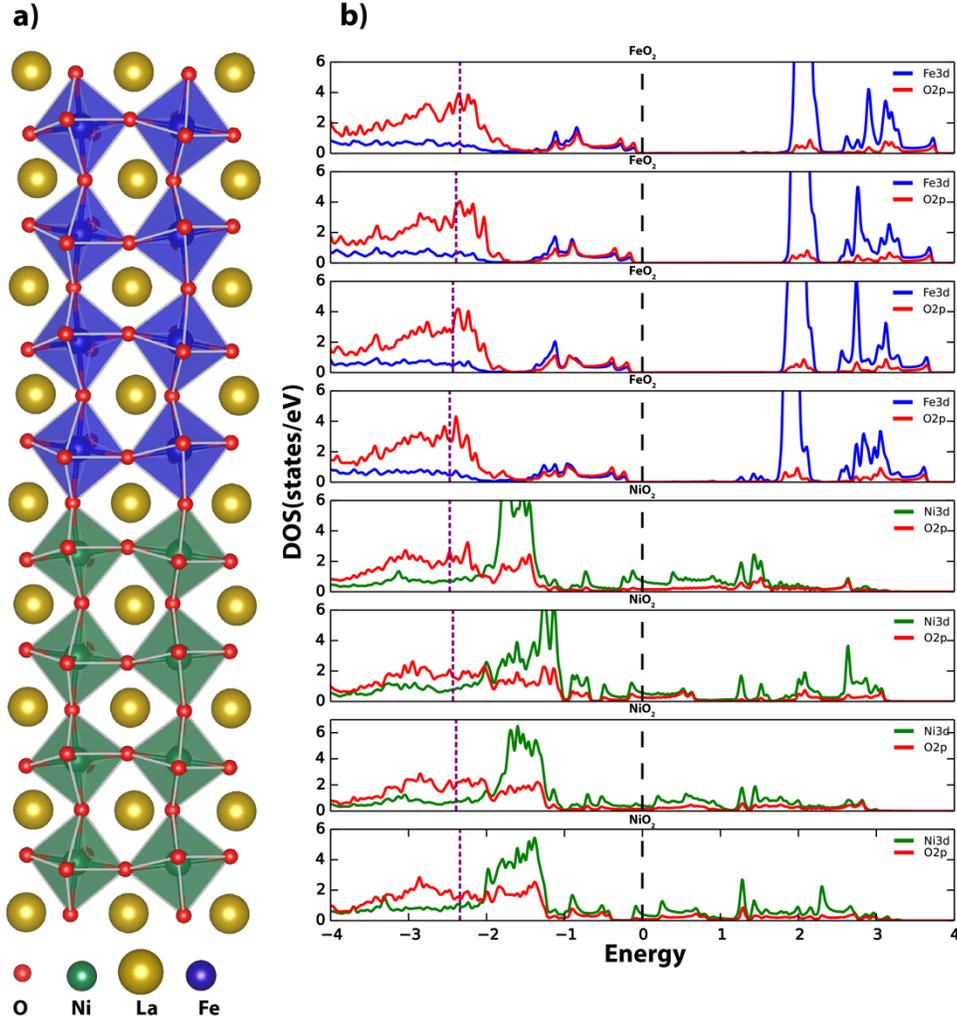

Figure 1: a) LFO/LNO superlattice model used for DFT band alignment predictions; b) Orbital projected density of states projected onto different $FeO_2$ and $NiO_2$ layers. Black dashed line is the Fermi energy and green dashed line represents the O 2p band center.

Table 1: Oxygen 2p band center in different LaO, $FeO_2$ and $NiO_2$ layers. Energy values are in eV.

|  | $E_p(FeO_2)$ | $E_p(LaO)_{LFO}$ | $E_p(LaO)_{LNO}$ | $E_p(NiO_2)$ |
|---|---|---|---|---|
| Interface 1 | -2.47 | -2.40 | -2.48 | -2.47 |
| Internal 1 | -2.43 | -2.31 | -2.27 | -2.43 |
| Internal 2 | -2.39 | -2.29 | -2.22 | -2.39 |
| Interface 2 | -2.34 | -2.22 | -2.22 | -2.34 |

*Experimental Band Alignment Studies*

To test the theoretical results determined from DFT, a series of films were grown via MBE and measured using *in situ* XPS [41,42]. This method is based on constant binding energy difference between a chosen core level and the valence band maximum (VBM)of a material. Kraut et. al. originally used gaussian broadened theoretical valence band density of states (VBDOS) to fit the experimental valence band data to determine the VBM [48]. In ternary complex oxides, accurate theoretical calculation of VBDOS is complicated and computationally expensive [14]. Chambers et. al proposed a rather simple method of estimating the VBM using extrapolation of the linear part of leading valence band edge to zero level background, which has been found to predict the VBM with sufficient accuracy [49].

Initial LFO and LNO films were grown separately and used as references for determination of the energy difference between core levels and the valence band maximum in each material. These results are shown in Figure 2. The LFO and LNO valence bands extracted from XPS are shown in Figure 2(a-b) and determination of VBM by extrapolating the linear region of leading valence band edge is shown in the inset. Linear extrapolation method yields VBM of LFO at -0.2 eV and the VBM value for LNO is -1.1eV. Note that due to the use of the electron flood gun and alignment of the O 1s peak to 530 eV, binding energies are not measured relative to the Fermi level.

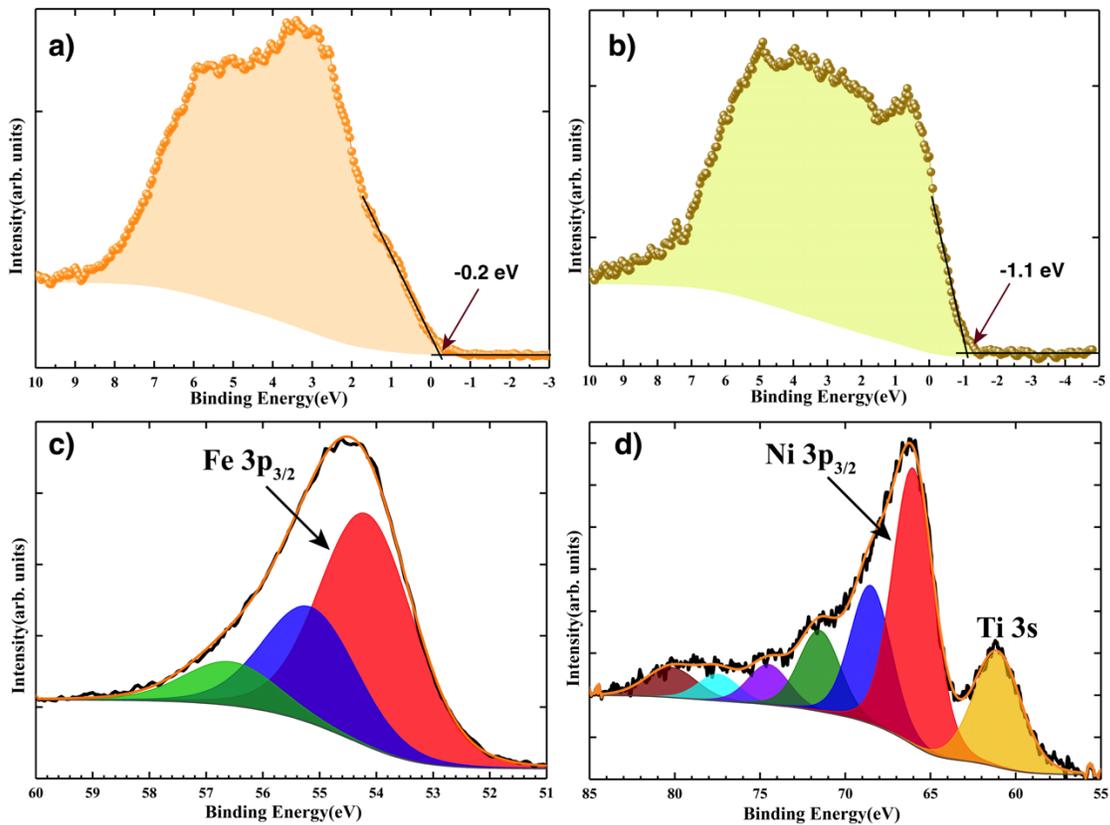

*Figure 2: Valence band XPS spectra of a) 6nm thick LFO sample b) 3nm LNO sample and fitting of valence band leading edge using linear extrapolation method  c) Fe 3p XPS region of LFO sample d) Ni 3p XPS region for LNO sample.*

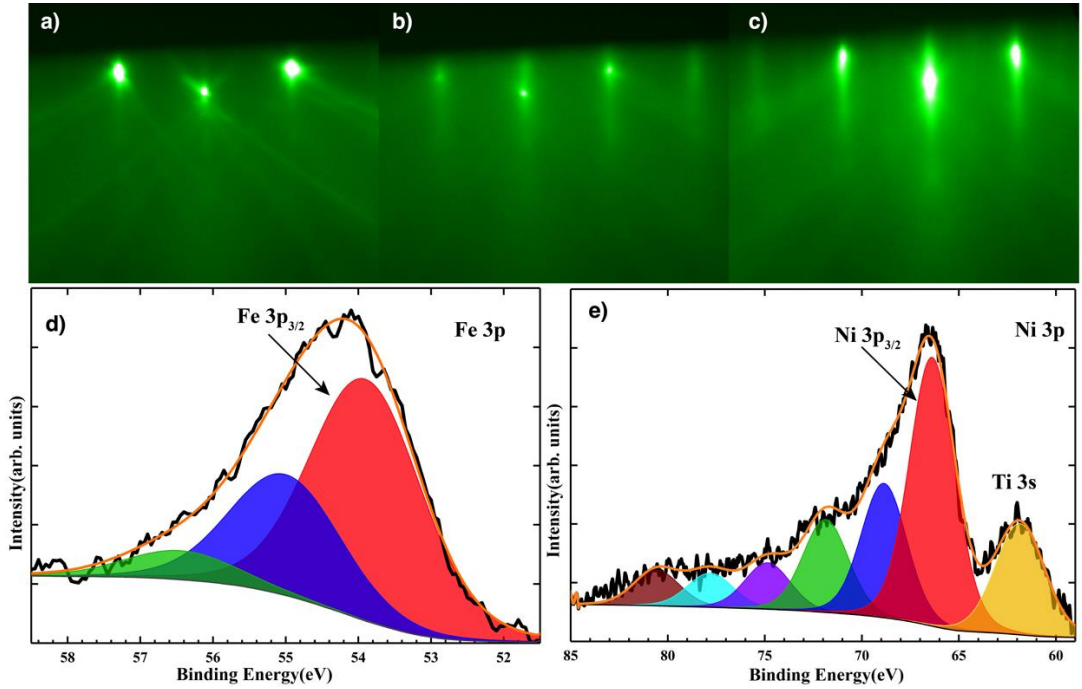

*Figure 3: RHEED images of a) Nb:STO substrate b) 4 unit cell LNO-NbSTO c) LFO-LNO-NbSTO with 2 unit cell LFO d)Fe 3p XPS region of LFO-LNO sample-1 e) Ni 3p XPS region of LFO-LNO sample-1*

The valence band offset is calculated as $\Delta E_V = (E_{CL} - E_V)_{LNO} - (E_{CL} - E_V)_{LFO} - \left(E_{CL}^{Ni\,3p} - E_{CL}^{Fe\,3p}\right)_{LFO/LNO}$, where $E_{CL}$ is the chosen core level energy and $E_v$ is the position of valence band maximum. The core level and VBM binding energy difference is measured in LFO and LNO thin films, and the core level energy difference is measured in the heterostructure. The individual LFO (~6nm) and LNO (~3nm) thin films used to determine the core level and VBM energy difference for band offset measurement. The Fe 3p and Ni 3p XPS regions are shown in Figure 3(c) and 3(d). Deconvolution of Fe 3p regions shows that there is additional satellite peak along with $3p_{1/2}$ and $3p_{3/2}$. Fe $3p_{3/2}$ was chosen as a core level in LFO. Similarly, deconvolution of Ni 3p reveals multiple peaks that must be deconvolved under a consistent standard for analysis across samples [50]. The lowest Ni $3p_{3/2}$ feature (shaded region in red) was chosen as the LNO core level for $E_{CL}^{LNO}$. In order to maintain consistency in measurement, full width at half maximum, area ratio, energy difference between various valence and spin multiplet peaks were constrained to be the same for all measurements. The core level binding energies, VBM and VBO are presented in the Table 2. The VBO for all three heterostructure samples are consistent with value of ~0.3eV. Considering the error bar this is in close agreement with the VBO estimated from first principles calculation.

*Table 2: Core level, VBM and band offset of LFO, LNO and LFO/LNO samples. The thickness of heterostructure samples: LFO/LNO-1 is 2uc/3uc, LFO/LNO-2 is 4uc/3uc and LFO/LNO-2 is 4uc/4uc. The energy values presented in the table are in eV.*

| Sample | Ni $3p^{3/2}$ | Fe $3p^{3/2}$ | VBM | $\Delta E_v$ |
|---|---|---|---|---|
| LFO | - | 54.17 | -0.2 | |
| LNO | 66.02 | - | -1.1 | |
| LFO/LNO-1 | 66.37 | 53.90 | -0.65 | 0.28(0.1) |
| LFO/LNO-2 | 66.48 | 54.07 | -0.70 | 0.33(0.1) |

*Electrocatalysis*

Interfacial band engineering with epitaxial films can have a significant impact on functional properties, including electrocatalysis. To examine this impact, we studied the OER electrocatalysis using the epitaxial LFO/LNO thin film samples above. Films were converted into electrodes through back contact with a commercial glassy carbon disk and assembled into a rotating-disk shaft electrode. **Figure 4** shows the anodic scan for cyclic voltammetry data collected for each film at a scan rate of 20 mV s$^{-1}$ while spinning the disk electrode at 2000 rpm to prevent bubble formation. Each voltammogram was collected following 25 continuous scans over the potential range 0.83 - 2.23 V vs RHE at 20 mV s$^{-1}$. The catalytic current densities observed at 1.6 V vs RHE ($\eta_{OER}$=370 mV) were found to be 26 µA cm$^{-2}$ for LFO/LNO-1 and 55 µA cm$^{-2}$ for LFO/LNO-2. The slightly larger current observed for LFO/LNO-2 may be due to the difference in thickness of the LFO layer. By comparison, the current density observed for a 5 u.c. LFO film (~2 nm) directly on an n-STO substrate (i.e. without an intermediate LNO layer) was recently reported by our groups to be 0.2 µA cm$^{-2}$ [15]. This data is reprinted in **Figure 4** for direct comparison with the present LFO/LNO films.

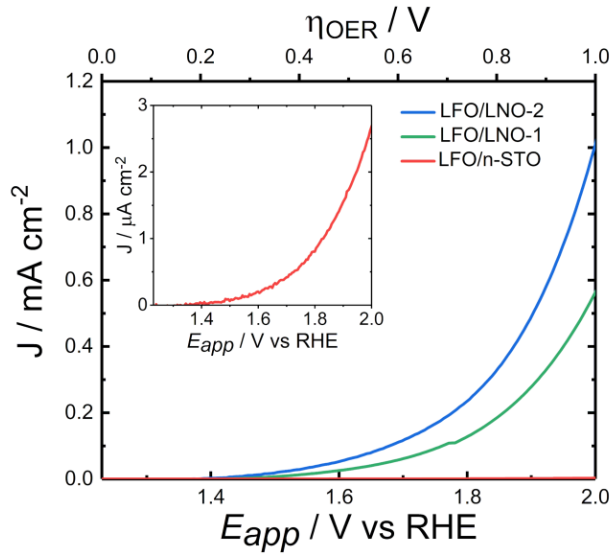

*Figure 4: Anodic scans obtained from CV for LFO/LNO/n-STO films. Data collected in $O_2$ saturated 0.1 M KOH aqueous electrolyte at 20 mV s$^{-1}$ scan rate and 2000 rpm rotation. (Inset) Current density measured for 2 nm thick LFO film deposited directly on n-STO (data taken from Burton et al [15])*

**Discussion**

The two samples exhibit a 130-fold increase (LFO/LNO-1) and 275-fold increase (LFO/LNO-2) in catalysis observed for the LFO/LNO films compared to LFO alone is remarkable. We attribute this result to the interfacial hole doping due to the smaller VBO observed for the LFO/LNO interface (0.28-0.33 eV) than the LFO/STO interface reported previously (2.2 eV). The approximate band alignment diagram based on our experimental and theoretical determination of VBO based on our previous work [15] is summarized in Figure 5. Electrons transferred from LFO leave behind holes which move towards the surface (electrolyte-LFO interface) and contribute to water oxidation in the same fashion as holes produced by dopants in, for example, (La,Sr)FeO$_3$ [30]. The hole effective mass is $0.61m_e$, which makes the transportation of these holes in external electric field easier. Thus, creation of holes at the LFO/LNO interface will increase the surface oxidation kinetics, resulting in higher OER activity.

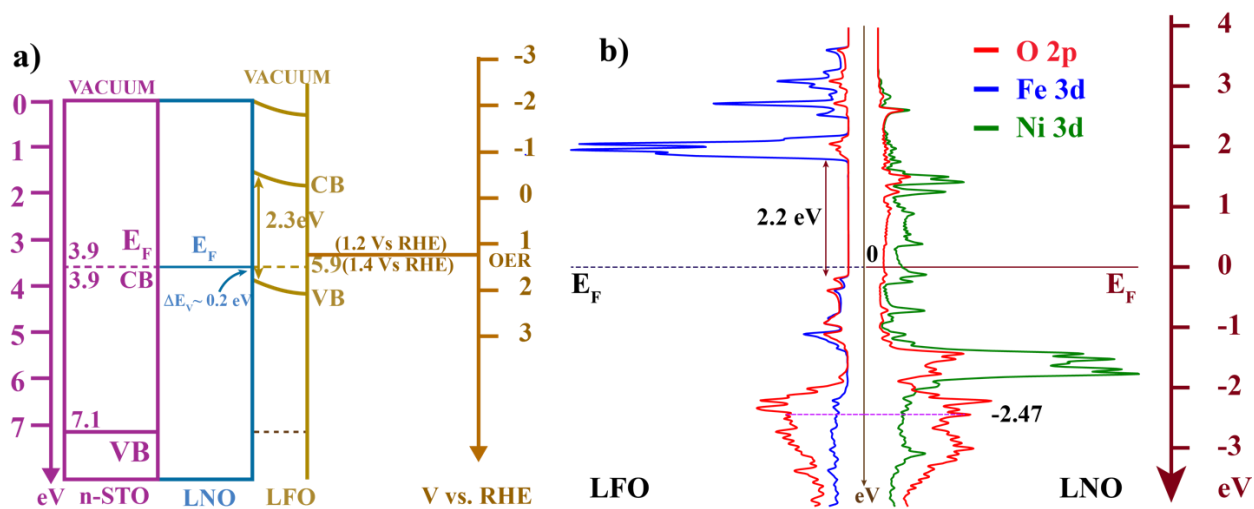

*Figure 5: a) Experimental band alignment at STO/LNO/LFO interface b) Theoretical band alignment at LNO/LFO interface. Dashed line(magenta)at -2.47eV represents the O 2p band center at interfacial FeO2 and NiO2 layers.*

We see that the VBM in LFO/LNO moves much closer to the OER activation energy (1.23V vs RHE) compared to the LFO/*n*-STO interface, reducing the overpotential significantly [15]. Moreover, during OER catalysis, electrons must travel from solution, through the valence band of LFO, and into the valence (conduction) band of LNO (*n*-STO). The band offset at the LFO/LNO(n-STO) interface thus represents an uphill barrier for charge transfer. The insertion of LNO between STO and LFO thus decreases this barrier significantly and results in greater electrocatalysis.

It is also noteworthy that we do not see evidence of significant formation of $Fe^{4+}$ formal charge states based on either the DFT predictions or the XPS analysis of the interfaces. Previous analysis by Wang et al. of alloy La(Fe,Ni)O$_3$ thin films had showed evidence of $Fe^{4+}$ and $Ni^{2+}$ formal charges in some samples with strongly enhanced electrocatalytic performance [24]. DFT models of these alloys also predicted formation of $Fe^{4-\delta}$ formal charge states at low Fe concentrations (12.5%) but did not use Hubbard U parameters to reproduce the experimental band gap of pure LFO [24]. While we do predict favorable hole transfer into LFO due to the small valence band offset between LNO and LFO, when accounting for the band gap of LFO in the model our analysis does not suggest that significant $Fe^{4+}$ formal charge would be expected. Further studies of LFO/LNO heterostructures using electron microscopy and X-ray absorption spectroscopy could help to address these open questions.

**Conclusions**

We have investigated the electronic structure of the epitaxial LFO/LNO interface theoretically and experimentally and related these insights to the strong OER activity of these heterostructures. Theoretical predictions indicate that the LFO VBM should lie ~0.2 eV below the Fermi level when pinned by an interface with LNO. Comparisons with experimental VBO measurements are in good agreement with these DFT predictions. We find that the OER activity of the heterostructure is much higher (~275 times for the best performing sample) when compared with the LFO and LNO thin films of same thicknesses. The small VBO offers low energy barriers for charge carriers and promotes charge transfer from LFO valence band to LNO. This minimal charge transfer results in partial oxidation of $Fe^{3+}$ and partial reduction of $Ni^{3+}$, which leads to high OER activity. Thus, our study highlights the importance of band engineering in developing a highly efficient OER catalyst.

**Acknowledgements**

R.P., A.R.B, B.H.F., and R.B.C. acknowledge support from the National Science Foundation (NSF) Division of Materials Research through grant NSF-DMR-1809847. Additionally, A.R.B. acknowledges support from the Alabama EPSCOR Graduate Research Scholars Program. M.A.K. acknowledges computational resources from the Hopper HPC system at Auburn and funding support from NSF through NSF-DMR-1848344.

# Band-Engineered LaFeO₃-LaNiO₃ Thin Film Interfaces for Electrocatalysis of Water

Rajendra Paudel[a], Andricus R. Burton[b], Marcelo A. Kuroda[a], Byron H. Farnum[b], and Ryan B. Comes[a]

[a]Department of Physics, Auburn University, Auburn, AL 36849 USA
[b]Department of Chemistry and Biochemistry, Auburn University, Auburn, AL 36849 USA

*Supplemental Information*

## Density Functional Theory Modeling

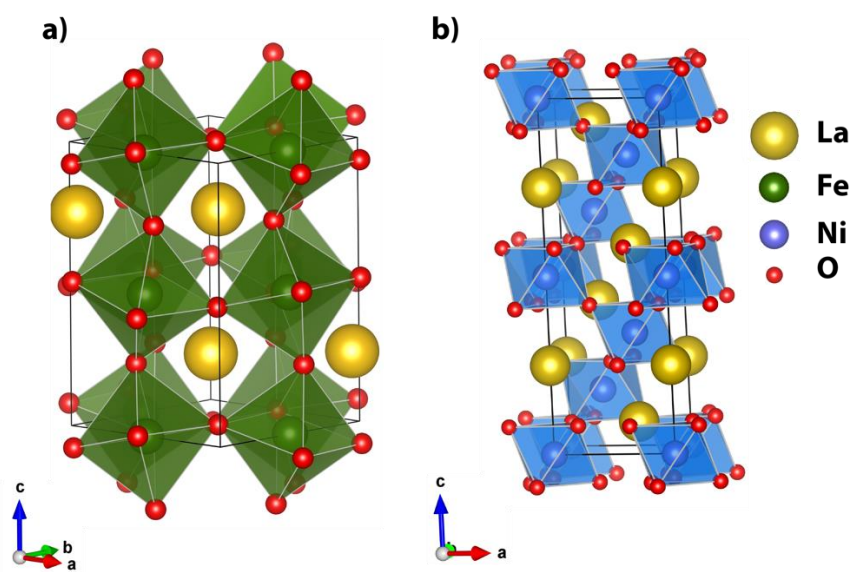

*Figure S1: a) Orthorhombic crystal structure of LaFeO₃; b) Rhombohedral crystal structure of LaNiO₃*

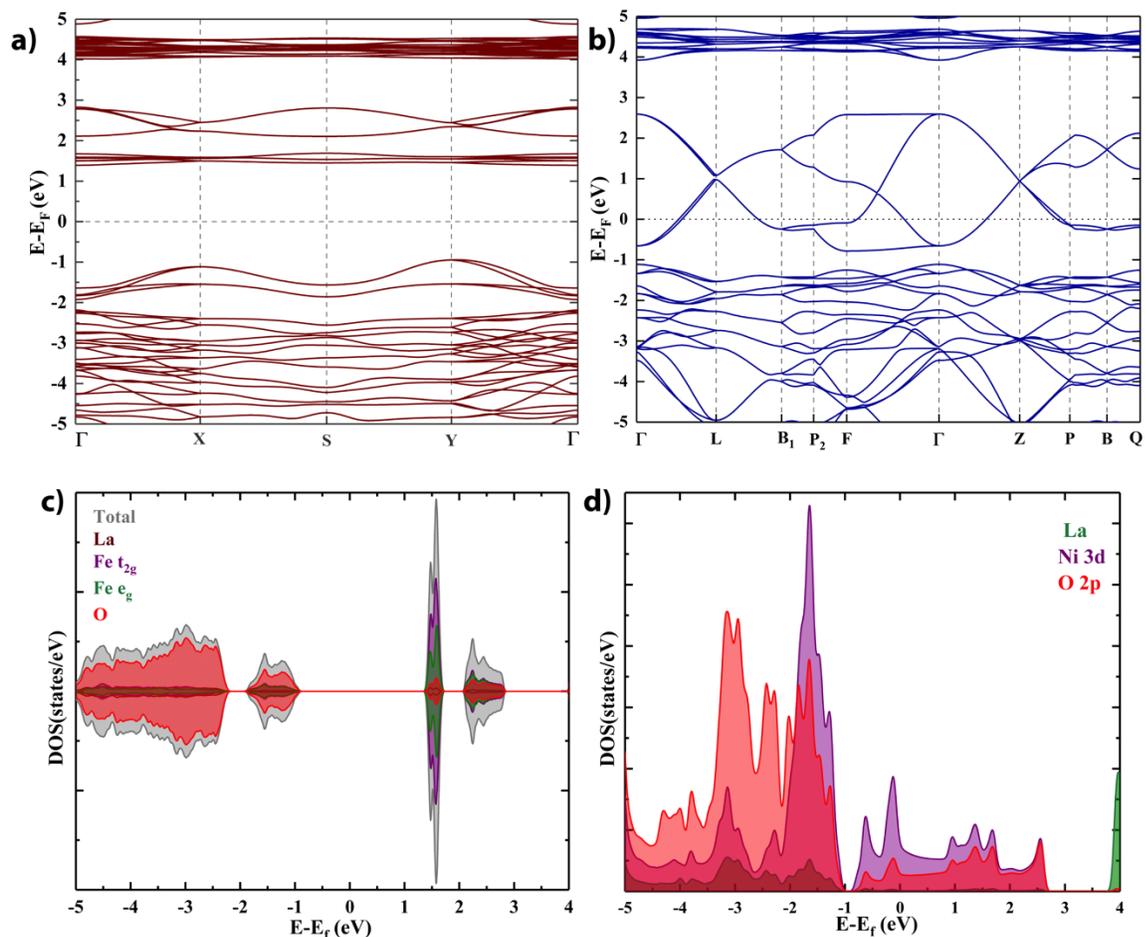

*Figure S2: a) Electronic band structure of LaFeO$_3$, b) Electronic band structure of LaNiO$_3$, c) Density of states of LaFeO$_3$ on various atoms and atomic orbitals, d) Density of states of LaNiO$_3$*

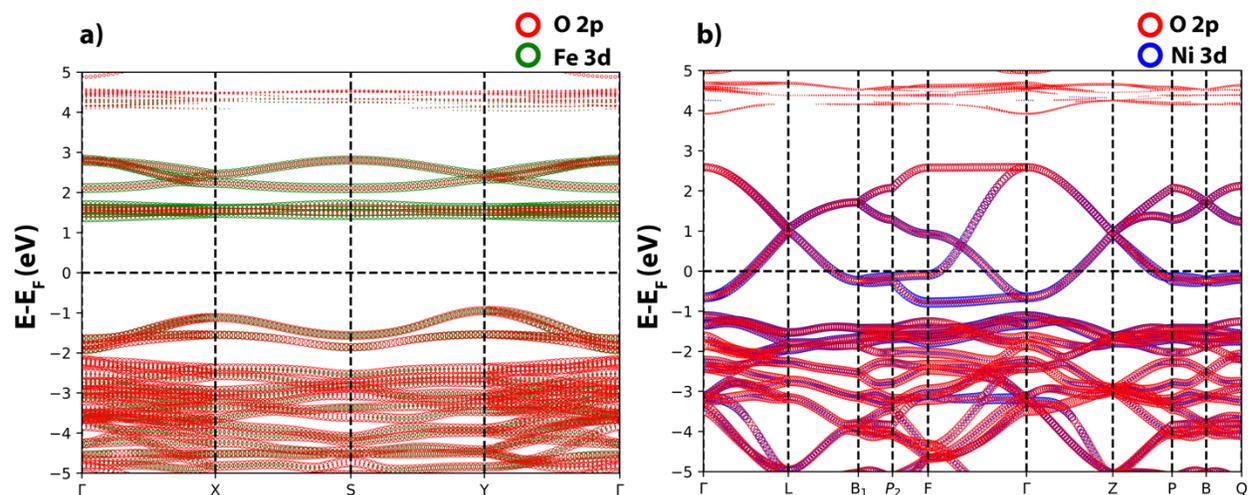

*Figure S3: a) Electronic band structure of LaFeO3 projected on different orbitals. b) Orbital projected electronic band structure of LaNiO3 . The area of the circle is proportional to the weight of the projection on specific orbital at given k- value.*

LFO at room temperature has a G-type antiferromagnetic phase with orthorhombic crystal structure. Occupied $e_g$ states of $Fe^{3+}$ lie in the valence band and unoccupied $t_{2g}$ forms the state close to the

Fermi level in conduction band. The valence band edge is mostly O 2p states hybridized with $e_g$ states. Spin polarized DFT+U calculation estimates the bandgap ~2.2eV, which is in close agreement with experimental band gap of 2.3eV [1]. LNO, on the other hand, is a rhombohedral paramagnetic metal at room temperature. Fig S2 shows the band structure and DOS of LNO. The bands near the fermi level are strongly hybridized Ni 3d and O2p orbitals, which is responsible for metallicity of LNO. $Ni^{3+}$ $t_{2g}$ states which are almost completely filled, forms the valence band and partially occupied $e_g$ states cross the fermi level forming metallic band. This is consistent with experimentally observed electronic structure of LNO [2].

**Hole Density Calculations**

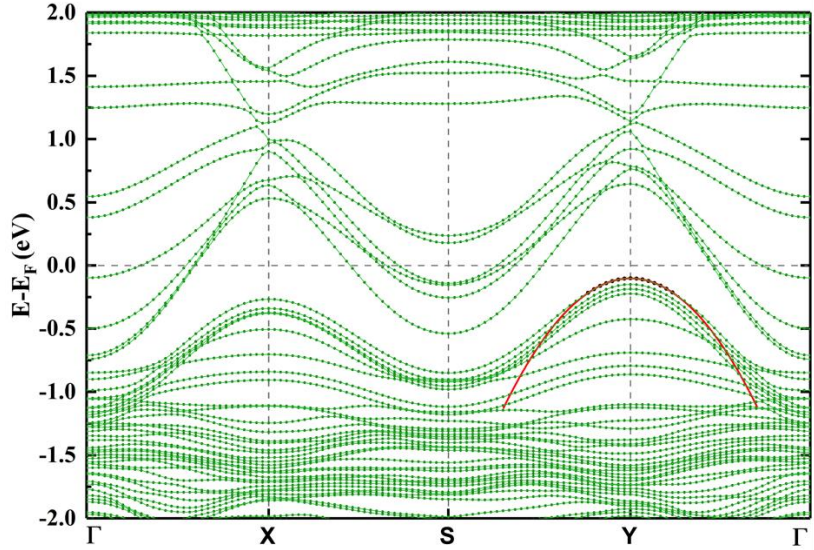

*Figure S4: Parabolic fit of the hybridized O 2p-Fe 3d electronic bands in the to calculate the hole effective mass at the VBM along (100) and (010) directions*

The energy of the electron for given *k*-value is
$$E(k) = E_0 + \frac{\hbar^2 \kappa^2}{2m}$$
And the effective mass of holes in the energy band is calculated from second derivative of the E versus *k* curve as,
$$\frac{1}{m*} = \frac{1}{\hbar^2}\left(\frac{\partial^2 E(k)}{\partial k^2}\right)$$

Considering the symmetric nature of the electronic bands we calculated the effective mass of hole along **S-Y-Γ** k-path. The parabolic fit to the bands form VBM is shown in Figure S4. The effective mass of holes found to be $0.6 m_e$.

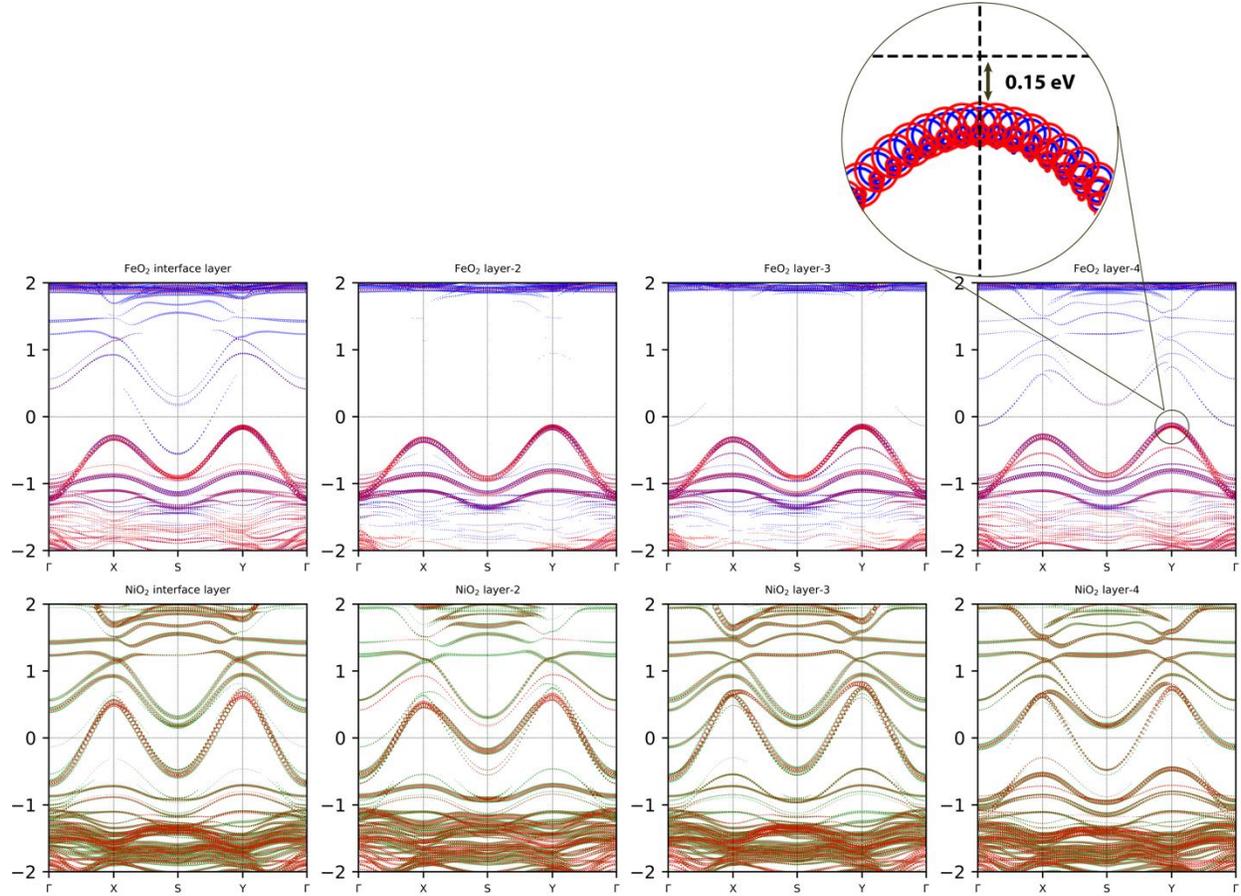

*Figure S5: Band structure of LaFeO$_3$-LaNiO$_3$ heterostructure projected on different atomic layer according to the orbital weight*

From the information about density of states, position of Fermi energy and valence band maximum we can calculate the number of thermally generated holes at temperature T as,

$$P_o = \int_{-\infty}^{E_v} g_v(E)[1 - f_F(E)]dE$$

$$\text{Where,} \quad 1 - f_E(E) = \frac{1}{1 + exp(\frac{E_F - E}{kT})}$$

For ($E_F$-$E_v$)$\gg$$k$T,

$$1 - f_E(E) \approx exp[\frac{-(E_F - E)}{kT}]$$

And density of holes is calculated as, $n = \frac{P_o}{unit\ cell\ Volume}$

The hole density calculated for different valence band offset is presented in the Table S1.

*Table S1: Density of holes on valence band for different band offset values*

| Band Offset(eV) | Hole density(cm$^{-3}$) |
|---|---|
| 0.0 | $6.81\times10^{19}$ |
| 0.1 | $1.43\times10^{18}$ |
| 0.2 | $3.01\times10^{16}$ |
| 0.3 | $6.35\times10^{14}$ |